# Computational analysis and performance optimization of $SrScCu_3Se_4$-based solar cells using COMSOL Multiphysics

Khalid Mahmud[1], Md. Nahid Hasan[1], Md. Islahur Rahman Ebon[2], Bipanko Kumar Mondal[3] and Jaker Hossain[1*]

[1]*Photonics & Advanced Materials Laboratory, Department of Electrical and Electronic Engineering, University of Rajshahi, Rajshahi 6205, Bangladesh.*

[2]*Department of Electrical & Electronic Engineering, Gono Bishwabidyalay, Savar, Dhaka 1344, Bangladesh.*

[3]*Department of Electrical and Electronic Engineering, Begum Rokeya University, Rangpur, Rangpur 5400, Bangladesh.*

**Abstract**

Transition metal-based quaternary semiconductors show strong magneto-optical and thermoelectric properties, yet their photovoltaic performance in realistic three-dimensional device architectures remains underexplored. In this work, we investigate a n-ZnSe/p-$SrScCu_3Se_4$/$p^+$-$WSe_2$ quaternary chalcogenide heterostructure using three-dimensional finite-element simulations in COMSOL Multiphysics. The model self-consistently couples wavelength-dependent optical generation, drift–diffusion carrier transport, and thermal loss analysis under AM1.5G, 1-sun illumination within a fully coupled opto–electro–thermal framework. The effect of absorber width, acceptor level, and bulk defects on photovoltaic performance, carrier generation, and recombination profile is investigated in this study. Quaternary chalcogenide device produces an open circuit voltage ($V_{OC}$) of 1.02 V, short-circuit current density ($J_{SC}$) of 32.472 mA/cm$^2$, fill factor (FF) of 88.212 %, and power conversion efficiency (PCE) of 29.217% under optimized conditions. The quantum efficiency (QE) results indicate that the device effectively transforms incident light into charge carriers within the visible spectrum, however absorption and carrier collecting performance diminish in the near-infrared range. The electrothermal simulations indicate modest, spatially non-uniform temperature increases relevant to Joule heating and nonradiative recombination heating within the active layers. Overall, these findings ensure that $SrScCu_3Se_4$, a quaternary chalcogenide, is a promising absorber material and offers a strong experimental design consideration for obtaining high-performance, thermally stable three-dimensional photovoltaic device topologies.

## 1. Introduction

The fast and mostly unmanageable expansion in population has significantly escalated worldwide energy demand, rendering energy security a critical concern for contemporary society. The growing scarcity of conventional energy resources poses serious constraints on economic development, technological progress, and environmental sustainability. To tackle these issues, considerable efforts are focused on the development and deployment of innovative inventions, as well as unique chemicals and physical systems, capable of efficiently generating, transforming, storing, and distributing energy in various types. These emerging solutions aim not only to meet the rising energy demand but also to ensure long-term reliability, efficiency, and sustainability of future energy infrastructures [1]. Solar energy offers a clean and sustainable solution to growing global energy demands, with solar cells converting abundant sunlight directly into electricity through the photovoltaic effect, making them key technologies for a greener future [2]. Solar technology enables power generation across both residential and commercial sectors and is widely applied in diverse fields, including spacecraft propulsion, water pumping systems, solar home installations, remote and off-grid facilities, communication infrastructures, satellites, reverse osmosis plants, and large-scale power generation in megawatt-level power plants [3]. Owing to its wide range of critical applications and its profound impact on modern society, solar technology has drawn sustained attention from the global research community. Since the pioneering observation of the photovoltaic effect by Edmond Becquerel in 1839, extensive research on photovoltaic cells has been continuously pursued. Over the years, researchers have focused on addressing key challenges to enable large-scale commercialization, including improving cost effectiveness, enhancing conversion efficiency, and ensuring long-term reliability and operational stability. These efforts aim to make solar energy harvesting not only technologically viable but also economically competitive and sustainable for widespread adoption [4-5].

Recent advances in photovoltaic research have primarily focused on improving device efficiency while concurrently lowering manufacturing costs to satisfy the requirements of large-scale industrial applications. Consistent with the Shockley–Queisser detailed balance limit, the theoretical maximum efficiency of dual-heterojunction solar cells is estimated to lie in the range of approximately 42% to 46% [6-8]. Martí and Luque have recently proposed, through theoretical analysis, that a three-terminal dual-heterojunction bipolar transistor–based solar cell could potentially attain an efficiency of approximately 54.7% [9].

Transition-metal-based quaternary semiconductors have attracted considerable research interest in recent years due to their diverse and promising application potential, particularly in photocatalysis, photovoltaic technologies, and the emergence of unconventional superconductivity in iron-based compounds [10-12]. Moreover, copper-based oxyselenide compounds exhibit significant magneto-optical responses and possess remarkable thermoelectric properties, further enhancing their relevance for advanced functional material applications. Among the materials, $SrScCu_3Se_4$ (SSCSe) has some marvelous characteristics. The strong light-absorption capability, effective utilization of solar irradiation, and robust structural and chemical stability of these materials make them highly suitable candidates for solar cell applications [13]. These materials exhibit a compelling combination of electrical, optical, elastic, and thermoelectric properties, whose synergistic interplay offers significant advantages for a wide range of technological and industrial applications [14]. A thorough understanding and effective utilization of the investigated properties are essential for advancing materials research and driving further scientific innovation. Such insights not only deepen fundamental knowledge of material behavior but also facilitate the development of next-generation technologies, enabling the design of novel devices and applications across diverse technological domains [15]. The investigated materials exhibit excellent structural stability, as confirmed by the absence of imaginary (negative) phonon frequencies, indicating dynamic stability. Analysis of the electronic structure reveals that the s and p orbitals of selenium contribute only marginally, whereas the Cu *d* states play a dominant role in shaping the valence band region. The calculated energy band gap of SSCSe is approximately 1.35 eV, placing it within a favorable range for optoelectronic and photovoltaic applications. Furthermore, SSCSe demonstrates relatively high hardness while also exhibiting notable compressibility, reflecting a balanced combination of mechanical strength and elastic flexibility that is advantageous for practical device integration [13].

The incorporation of a back surface field (BSF) layer plays a significant role in enhancing photon management within solar cell devices by promoting a more uniform distribution of incident photons throughout the absorber layer and mitigating spatial non-uniformities in light absorption. In addition to its electrical benefits, the BSF layer contributes to improved optical performance by reflecting unabsorbed photons back into the active region, thereby increasing the effective optical path length and overall absorption probability. Owing to these advantages, BSF thin films are currently extensively produced and incorporated as effective light-trapping layers in several solar technologies, including c-Si, III–V compound, and CdTe-based photovoltaic devices. More recently, BSF layers have also been explored as viable alternatives to conventional antireflection coating (ARC) layers. A key advantage of employing a BSF layer in place of a traditional ARC is its dual functionality: while minimizing optical losses at the rear interface, it simultaneously enhances light harvesting and carrier collection efficiency, leading to improved photovoltaic performance and device reliability [16]. Furthermore, an appropriately engineered BSF layer is essential between the back contact and the photo absorber to effectively suppress carrier recombination [17]. A viable option for use as a BSF layer is tungsten selenide ($WSe_2$), which is a member of VI chalcogenide family with a lattice constant of 0.61 nanometers. It possesses an energy gap of about 1.7 eV. In addition, $WSe_2$ exhibits favorable physical and optical properties, including a high melting point (1473 K), chemical stability, high refractive index (~5.5 at 0.5 μm), and dielectric permittivity of 13.8, making it well suited for integration with CZTS-based solar cells as an effective BSF layer [18].

Furthermore, zinc selenide (ZnSe) has demonstrated considerable potential as an efficient window layer for photovoltaic applications owing to its wide optical bandgap of approximately 2.7 eV, which enables high transparency across a broad spectral range. In addition, the favorable electron and hole mobilities of ZnSe further contribute to its suitability for enhancing charge transport and overall device performance in photovoltaic systems [19]. ZnSe can be fabricated using various established techniques, including molecular beam epitaxy, thermal evaporation, metal–organic chemical vapor deposition, and electrochemical deposition. Additionally, its low sensitivity to humidity and nontoxic nature makes ZnSe particularly suitable for large-scale and industrial applications [20].

This work presents a novel n-ZnSe/p-SSCSe/$p^+$-$WSe_2$ heterostructure engineered for application as a high-performance solar cell. A comprehensive numerical investigation is carried out using the widely adopted COMSOL Multiphysics simulation platform to evaluate the device behavior under realistic operating conditions. The study systematically examines the influence of key structural and material parameters, including individual layer thicknesses, defects level, and carrier density, on the overall cell performance. In addition, the effects of interface defect states, operating temperature, series resistance, and shunt resistance are rigorously analyzed to provide a detailed understanding of their roles in governing carrier transport, recombination mechanisms, and power conversion efficiency. The simulation outcomes reveal promising insights into the optimization of device architecture and material properties, thereby highlighting a promising route for enhancing the performance and technological viability of SSCSe-based solar device.

## 2. Structure of the device and computation

### 2.1 Structure of the proposed quaternary chalcogenide photovoltaic device

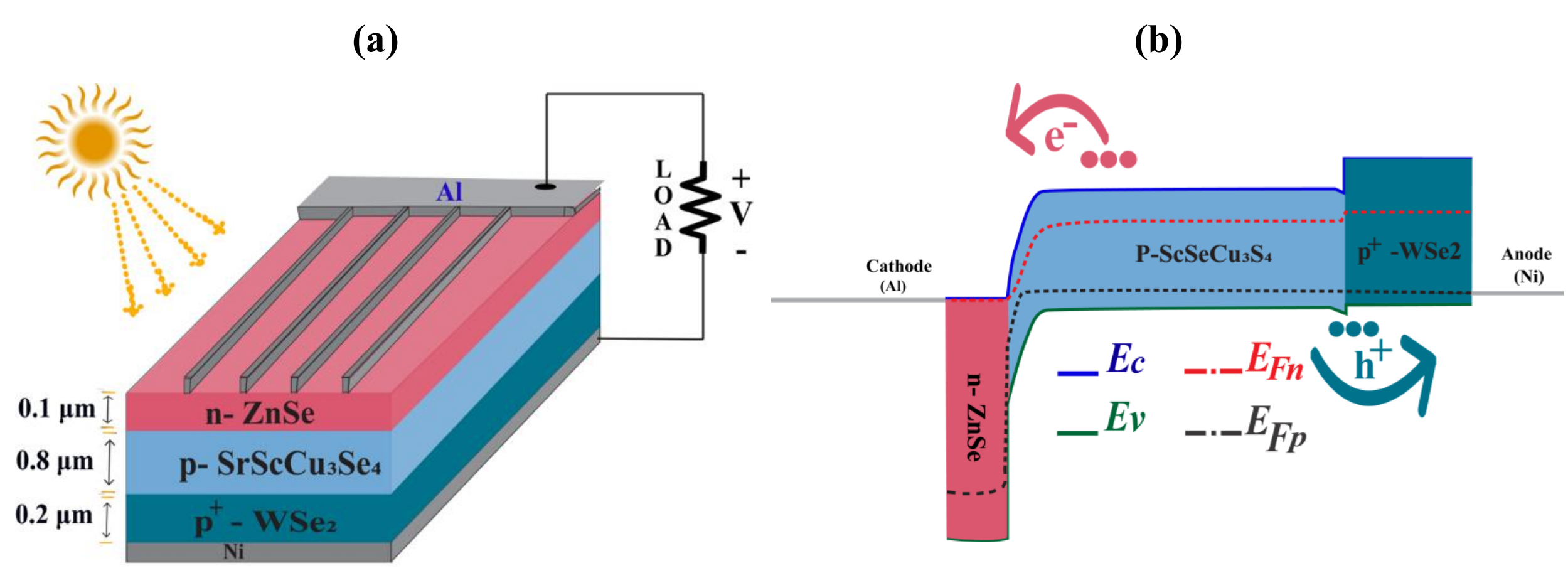

**Fig. 1:** (a) Schematic device configuration and (b) illuminated energy band diagram of the simulated n-ZnSe/p-$SrScCu_3Se_4$/$p^+$-$WSe_2$ heterojunction quaternary chalcogenide solar cell.

The conceptual device structure is shown in Fig. 1(a). The simulated transition-metal-based quaternary solar cell employs ZnSe as the electron transport layer (ETL), SSCSe as the absorber, and $WSe_2$ as the hole transport layer (HTL), with Al and Ni serving as the front and back contacts, respectively. Device performance is strongly governed by the quality of the ETL/absorber and HTL/absorber interfaces because these junctions control carrier selectivity, charge extraction, and interfacial recombination. Well-engineered interfaces promote efficient electron and hole collection while suppressing minority-carrier losses, which directly improves the overall power conversion efficiency.

Fig. 1(b) summarizes the energy band alignment of the n-ZnSe/p-$SrScCu_3Se_4$/$p^+$-$WSe_2$ device stack, where the relative positions of the conduction band edge and valence band edge control whether each interface favors carrier extraction or carrier blocking. Using the simulated material parameters and an electron-affinity-based band placement, ZnSe with Eg = 2.7 eV and $\chi$ = 4.09 eV forms a near-aligned conduction band with SSCSe ($E_g$ = 1.35 eV, $\chi$ = 4.10 eV), promotes low-resistance electron extraction into the ETL. At the same time, the significantly deeper valence band of ZnSe acts as an effective hole-blocking barrier, which limits hole leakage toward the ETL and suppresses interfacial recombination at the ETL/absorber junction [21]. At the absorber/HTL side, $WSe_2$ (Eg = 1.65 eV, $\chi$ = 3.70 eV) places its conduction band higher than that of the absorber, which confines electron leakage into the HTL, and retains its valence band close to the absorber valence band, permitting efficient hole extraction. This selective alignment is consistent with engaging Al (work function 4.2 eV) as the electron-collecting contact on the ZnSe side and Ni (work function 5.3 eV) as the hole-collecting contact on the $WSe_2$ side, as designated in Fig. 2(b). The physical parameters of the proposed device were derived from the literature and are summarized in Table 1.

**Table 1:** Parameters that were used for 3D simulation of SSCSe solar cell.

| Parameters | n-ZnSe | p-$SrScCu_3Se_4$ | $p^+$-$WSe_2$ |
|---|---|---|---|
| Depth (µm) | 0.1 | 0.8 | 0.2 |
| Energy gap $E_g$ (eV) | 2.7 | 1.35 | 1.65 |
| Affinity of Electron, $\chi$ (eV) | 4.09 | 4.1 | 3.7 |
| Dielectric Permittivity (relative) | 10 | 6.43 | 13.8 |
| CB effective DOS ($cm^{-3}$) | $1.8\times 10^{19}$ | $2.27\times10^{19}$ | $8.3\times 10^{18}$ |
| VB effective DOS ($cm^{-3}$) | $1.5\times 10^{18}$ | $3.92\times 10^{18}$ | $1.6\times 10^{19}$ |
| Electron mobility ($cm^2/(V.s)$) | 50 | 18.8 | 100 |
| Hole mobility ($cm^2/(V.s)$) | 20 | 60.64 | 500 |
| Donor concentration, $N_D$ ($cm^{-3}$) | $10^{18}$ | 0 | 0 |
| Acceptor concentration, $N_A$ ($cm^{-3}$) | 0 | $10^{17}$ | $10^{18}$ |
| Defect concentration ($cm^{-3}$) | $10^{14}$ | $10^{14}$ | $10^{14}$ |

For the SSCSe absorber, the carrier effective masses were extracted from the curvature of the calculated band dispersion E(k) using a parabolic approximation around the relevant band extrema. In this approach, the electron (hole) effective mass near the conduction-band minimum (valence-band maximum) is obtained from the second derivative of the fitted E(k) relation [22].

$$m^* = \frac{\hbar^2}{d^2E/dk^2} \tag{1}$$

The carrier mobility was then estimated using the relaxation-time expression [23].

$$\mu = \frac{q\tau}{m^*} \tag{2}$$

where q is the elementary charge and τ is the momentum relaxation time. Because τ is not directly available from the present electronic-structure calculation, a representative value of τ was taken as $10^{-14}$s was adopted to provide an order-of-magnitude estimate of the electron and hole mobilities. The effective densities of states at 300 K were calculated from the effective masses using the standard non-degenerate expressions [24]

$$N_c = 2\left(\frac{m_e^*\mathrm{KT}}{2\pi\hbar^2}\right)^{\frac{3}{2}} \tag{3}$$

$$N_v = 2\left(\frac{m_h^*\mathrm{KT}}{2\pi\hbar^2}\right)^{\frac{3}{2}} \tag{4}$$

Here, $m_h^*$, $m_e^*$, T, $K_B$, and $\hbar$ signify the effective masses of holes, electrons, the temperature, Boltzmann constant, and reduced Planck constant, in turn. The resulting $N_C$ and $N_V$ values quantify the number of available electronic states near the band edges and are useful for assessing carrier statistics and transport behavior in SSCSe. In particular, the relative magnitudes of $N_C$ and $N_V$ influence the balance between electron and hole populations and should be interpreted together with the dominant defect chemistry or experimentally inferred conductivity type.

### 2.2. 3D Simulation Framework

Computational modeling was carried out to clarify the governing physical mechanisms of the proposed solar cell and to estimate its attainable performance under standardized illumination. Device simulations were performed in the COMSOL Multiphysics 3D environment using the Semiconductor Module. In this framework, optical absorption and the resulting electron–hole pair generation are coupled to carrier transport, recombination, and charge collection at the electrodes. The photovoltaic response is obtained by solving Poisson's equation self-consistently with the steady-state drift–diffusion transport and continuity equations for electrons and holes, thereby determining the spatial distributions of electrostatic potential (Φ), electron concentration (n), and hole concentration (p). The governing equations are written as:

$$\nabla.(\varepsilon_0.\varepsilon_r.\nabla\Phi) = -\rho \quad (5)$$

where the space-charge density is $\rho = q(n - p + N_A - N_D)$. Here, q is the elementary charge, ε0 is the vacuum permittivity, εr is the relative permittivity, and $N_A$ and $N_D$ represent the densities of ionized acceptors and donors (including intentional dopants and fixed ionized impurities). Carrier conservation is enforced through the continuity relations,

$$\frac{\delta n}{\delta t} = \frac{1}{q}\nabla j_n + G_n - U_n \quad (6)$$

$$\frac{\delta p}{\delta t} = \frac{1}{q}\nabla j_p + G_p - U_p \quad (7)$$

where $G_n$ and $G_p$ denote the electron and hole generation rates and $U_n$ and $U_p$ denote the corresponding recombination rates. The electron and hole current densities include drift under the electric field and diffusion due to concentration gradients, given by

$$J_n = -q\mu_n n\nabla\Phi + qD_n\nabla n \quad (8)$$

$$J_p = -q\mu_p p\nabla\Phi + qD_p\nabla p \quad (9)$$

where μn and μp are the mobilities and Dn and Dp are the diffusion coefficients. The diffusion coefficients are linked to mobility through the Einstein relations,

$$D_n = \mu_n \frac{K_B T}{q} \quad (10)$$

$$D_p = \mu_p \frac{K_B T}{q} \quad (11)$$

where $k_B$ the Boltzmann constant and T the absolute temperature. The total current density is J = $J_n + J_p$. Trap-assisted Shockley–Read–Hall recombination is incorporated through $U_n$ and $U_p$ using user-defined material parameters in the Semiconductor Module, and steady-state operation is imposed by setting the time-derivative terms in the continuity equations to zero. By solving this coupled system under illumination, the current–voltage characteristics, power conversion efficiency, and spectral QE can be predicted. The incident illumination was modeled using the AM1.5G reference spectrum as a plane-wave input over the relevant tabulated wavelength range.

The optical generation profile was computed using a conventional physics-based approach and subsequently coupled to the electrical solver in COMSOL. First, the wavelength-dependent generation within each semiconductor layer was evaluated to construct the depth-resolved photogeneration rate. Absorption near the band edge was represented using a direct-transition

Tauc-type model [25-26]. The spectral generation was then integrated over the AM1.5G irradiation range to obtain the total depth-dependent generation function, $G_{tot}(z)$. This profile was finally imported into the Semiconductor Module as the carrier generation term, applied equally to electrons and holes, so that the electrical solution self-consistently reflects the optically generated carrier distribution. This opto-electrical coupling strategy follows the methodology reported in the referenced study [27].

To evaluate the spatial distribution of thermal losses across the device, volumetric heat-source outputs were extracted from the COMSOL Semiconductor Module based on the self-consistent drift–diffusion solution. In this study, thermal dissipation was attributed primarily to nonradiative recombination heating and Joule heating within the semiconductor stack. The recombination-related heat release was obtained from the locally resolved trap-assisted (Shockley–Read–Hall) and Auger recombination rates, while Joule heating was determined from the local electrical transport response within the layered structure. These quantities were post-processed to generate three-dimensional heat maps under short-circuit and applied-bias conditions, as discussed in Section 3.7. This spatial thermal-loss mapping procedure follows the methodology reported in the referenced study [27].

### 2.3 Strategy for meshing

A user-controlled mesh was generated for the full device geometry (Fig. 2). Each domain was meshed individually using a minimum and maximum element size of 2.24 nm and 52.2 nm, respectively. The maximum element growth rate was set to 1.35, and the curvature factor was set to 0.3 to maintain a smooth size transition while adequately resolving curved features. Swept meshing was applied across the layered stack, and the element distribution in the sweep direction was specified separately for each region. The absorber layer was discretized using 50 elements with an element ratio of 5, the window layer used 10 elements with an element ratio of 3, and the BSF layer used 10 elements with an element ratio of 5. A symmetric distribution was selected to ensure a consistent meshing pattern across the corresponding faces of the SSCSe absorber.

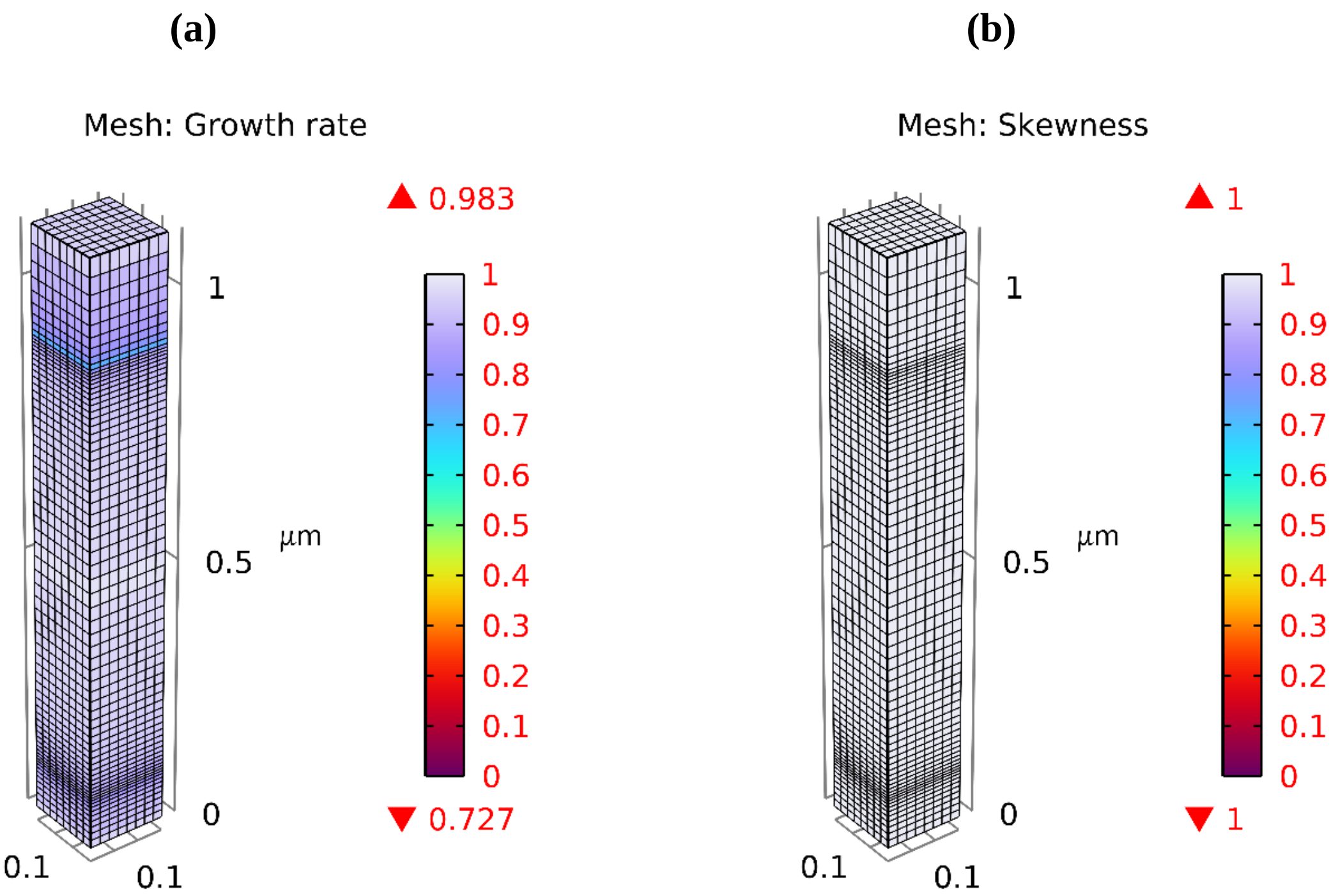

**Fig. 2:** The mesh quality created for (a) growth rate and (b) skewness.

Mesh quality was assessed using COMSOL's quality metrics based on skewness and element growth rate. Skewness evaluates angular distortion, penalizing elements that deviate from an ideal equiangular shape, with values approaching 1 indicating higher-quality elements and values approaching 0 indicating degenerated elements [26]. The growth-rate measure evaluates the local change in element size relative to neighboring elements, providing an additional check on mesh smoothness. Based on these criteria, the mesh quality was considered acceptable for the present analysis.

## 3. Results and discussion

### 3.1 Generation profile

The depth-dependent overall photogeneration rate, $G_{tot}(z)$, for the SSCSe-based device was evaluated, where the three-dimensional device structure includes a ZnSe window layer, a SSCSe absorber, and a $WSe_2$ back BSF layer. The three-dimensional spatial distributions of photogeneration rate, $G_{tot,}$ are presented in Fig. 3.

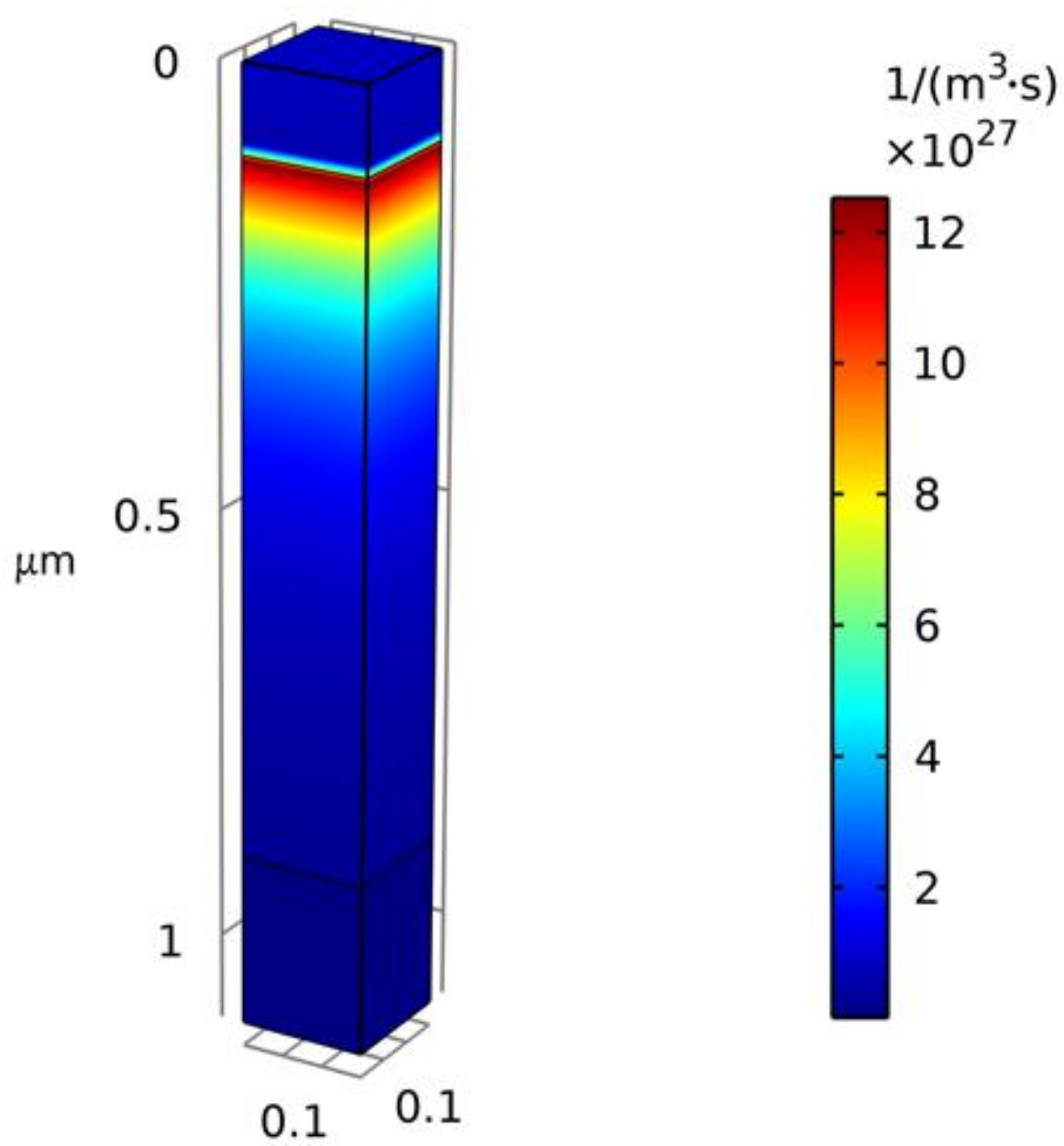

**Fig. 3:** 3D Photogeneration profile of SSCSe-based PV device.

Photogeneration is naturally highest close to the window/absorber interface, because the incident photon flux arrives at the absorber region through the ZnSe/SSCSe junction. The maximum photogeneration peak reaches 1.25 × $10^{28}$ $cm^{-3}$ $s^{-1}$ at the ZnSe/SSCSe junction. The bandgap of the SSCSe absorber is 1.35 eV, which can absorb maximum photons with energy larger than this. optical intensity attenuates with the absorber thickness, and photogeneration steadily declines as it moves toward the back surface. The generation rate at the absorber/BSF junction was found to be 3.95 × $10^{25}$ $cm^{-3}$ $s^{-1}$, which is much less than the window/absorber junction. The BSF layer provides a minor influence on the total photogeneration rate.

### 3.2. *Impact of Absorber layer on PV performance of* the $SrScCu_3Se_4$ device

Fig. 4 illustrates how the performance of a solar cell changes with variations in the absorber layer parameter. To evaluate the optimal values for each parameter, the thickness of the absorber layer, carrier concentration, and defect density in the SSCSe-based solar cell were thoroughly varied from 0.2 µm to 1.2 µm, $10^{15}$ $cm^{-3}$ to $10^{18}$ $cm^{-3}$, and $10^{13}$ $cm^{-3}$ to $10^{18}$ $cm^{-3}$, respectively whereas the variables of window and BSF layer are reserved constant as shown in Table 1.

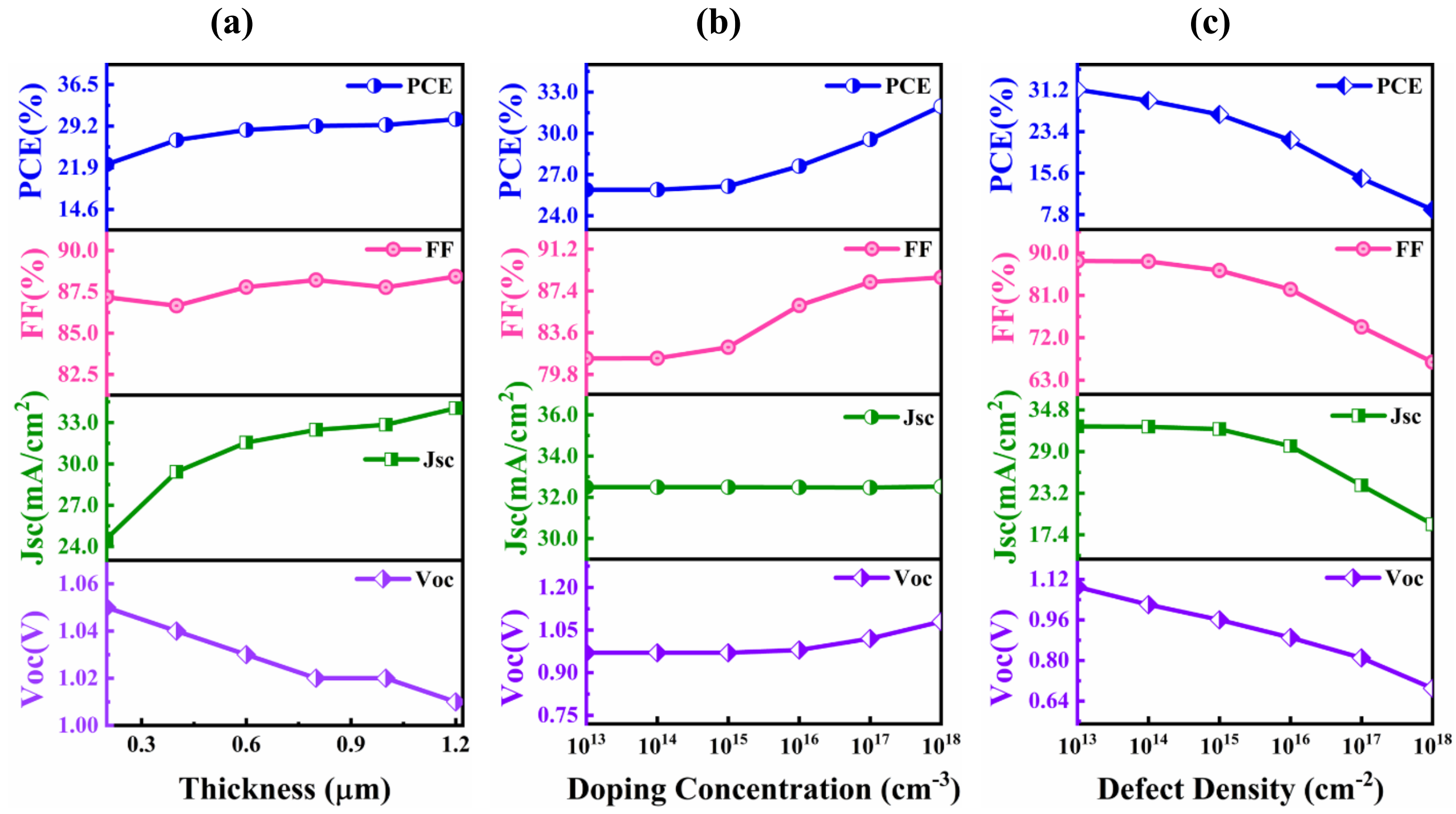

**Fig. 4:** Optimization of the performance parameters of SSCSe-based photonic device by varying (a) thickness (b) doping concentration and (c) defect density of the absorber layer.

Fig.4(a) indicates the device shows its maximum $V_{OC}$ = 1.05 V, when the photon-absorbing layer thickness is 0.2 μm and $J_{SC}$ of 34.042 mA/cm$^2$ at a thickness of 1.2μm. $V_{OC}$ tends to reduce as thickness increases, whereas JSC enhances substantially. As the thickness varied from 0.2 μm to 1.2 μm, the open-circuit voltage decreased from 1.05 V to 1.01 V, where $J_{SC}$ shows an upward trend from 24.572 mA/cm$^2$ to 34.042 mA/cm$^2$. Hence, increasing the thickness of the absorber layer improves the photon absorption [28]. Subsequently, the rise in $J_{SC}$ has elevated the PCE from 22.492% to 30.405% within the examined range. [29]. As well, the fill factor (FF) rises from around 87.176% to 88.413%, signifying a slight enhancement. However, $V_{OC}$ preserves falling, and JSC tends to saturate as the SSCSe $_{absorber}$ gets thicker. A thicker absorber raises the devices short-circuit current and decreases $V_{OC}$. At larger thickness values, recombination initiates to outweigh photon absorption, as shown by the saturation of $J_{SC}$ [30]. Thus, 0.8 μm is fixed as the absorber thickness for the following computations.

As the doping concentration in the SSCSe absorber is increased from $10^{13}$ to $10^{18}$ cm$^{-3}$, Fig. 4(b) displays that the device maintains a closely constant current density while $V_{OC}$ gradually enhances. At the concentration $10^{13}$ cm$^{-3}$, $V_{OC}$ equals 0.97 V, and it gradually increases to 1.08 V at $10^{18}$ cm$^{-}$

$^{3}$, most likely as an effect of large built-in potential and reduced diode ideality [29,31]. FF rises steadily and at higher doping it tends to saturation. Furthermore, PCE increases progressively, reaching its maximum performance (31.965%) at $10^{18}$ $cm^{-3}$. Doping concentration induced adjustments in recombination and ideality behavior, which optimize FF and PCE and are responsible for the inclusive enhancement [29,32]. Thus, the optimal doping concentration level for consequent computations is determined as $10^{17}$ $cm^{-3}$.

Fig. 4(c) demonstrates how a significant decrease in all parameters: $V_{OC}$, $J_{SC}$, FF, and PCE is triggered by an increase in defect density. For defect densities between $10^{13}$ $cm^{-3}$ and $10^{15}$ $cm^{-3}$, JSC is almost constant around 32.245 mA/$cm^2$, and that can be described by comparatively higher diffusion length and carrier lifetime, thereby decreasing recombination losses [29]. Whenever defect density tends to increase beyond this range, there is a sharp drop in JSC, which wanes the performance of the device. Furthermore, as defect density reaches from $10^{13}$ $cm^{-3}$ to $10^{18}$, FF decreases from 88.31% to 66.82% respectively. This phenomenon is specifically caused deo to ideality factor, and that has also been perceived in prior study [33]. Eventually, PCE reduces from 31.292% to 8.693% over the observed defect range, typically due to improved Shockley–Read–Hall recombination in larger defect density conditions [34].

### 3.3 *Impact of Window layer on PV performance of* the SSCSe device

To invent the optimal values for each parameter, the window layer thickness, carrier concentration, and defect density are sensibly varied within 0.05–0.3 µm, $10^{13}$-$10^{18}$ $cm^{-3}$, and $10^{13}$-$10^{18}$ $cm^{-3}$, respectively while all other parameters keep on unchanged (Table 1). Fig. 5 shows the subsequent variations in optoelectronic performance.

Fig. 5(a) illustrates the effect of the thickness variation of ZnSe on device performance. Because of the aptly large minority carrier lifetime and the relation of diffusion length to the ZnSe thickness, which results in slight recombination within the window layer, only minor changes are seen in $V_{OC}$, $J_{SC}$, FF, and PCE [35]. However, if the window layer is thickening too much, parasitic absorption could be introduced, which results in lower photovoltaic output [36]. As a consequence, $V_{OC}$ = 1.02 V, JSC = 29.21 mA/$cm^2$, FF = 88.21%, and PCE = 32.47% are found when the ZnSe thickness is optimized at 0.1 µm.

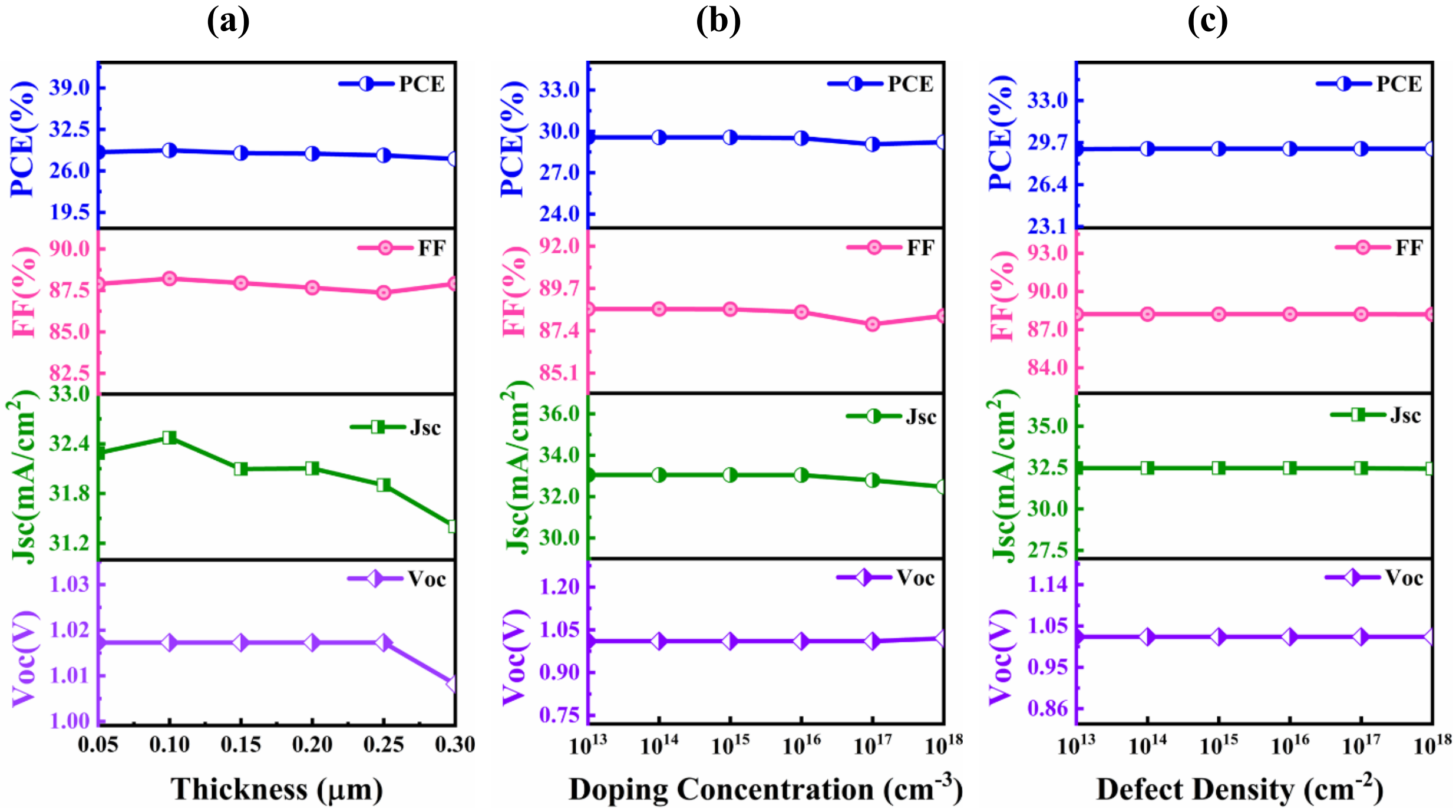

**Fig. 5:** Optimization of the performance parameters of SSCSe-based photonic device by changing (a) width (b) donor density and (c) bulk defects of the window.

The impression of varying the ZnSe layer's carrier concentration is exposed in Fig. 5(b), and a very negligible effect is observed on the solar cell's photovoltaic performance. It's evident that neither $V_{OC}$ nor PCE meaningfully changes when the concentration of ZnSe doping is enlarged. However, FF and JCS tend to decrease slightly at higher doping levels, and this can be described by the fact that higher doping concentrations induce recombination, particularly Shockley Read Hall (SRH) recombination, which has an impact on device performance, mostly JSC [37]. Furthermore, the enhanced recombination results rise in effective series resistance, which eventually lowers FF [38]. Additionally, Fig. 5(c) evidences that PV performance is nearly independent of defect density. Beyond this range, defect levels significantly increase recombination losses, shorten carrier diffusion length, and delay carrier transport, and those can ultimately result in a deterioration in the performance of the solar cell [29]. To increase device efficiency, it is therefore essential to carefully optimize the window-layer thickness, doping concentration, and defect density. To lessen recombination and uphold a suitable linking with the absorber at minimum doping conditions, the optimal defect density is observed at $10^{14}$ $cm^{-3}$.

### 3.4 *Impact of BSF layer on PV performance of* the $SrScCu_3Se_4$ device

How the $WSe_2$ back-surface field (BSF) layer interfere the n-ZnSe/p-SSCSe/$p^+$-$WSe_2$ double-heterojunction (DH) solar cell, which also acts as a bottommost absorber layer, is meticulously studied in this section. Fig. 6 shows that photovoltaic (PV) performance is less affected by variations of the parameters of the $WSe_2$ BSF layer, such as thickness, acceptor concentration, and defect density. It is obvious that as the thickness of the BSF layer is increased from 0.1 to 0.35 µm, the $J_{SC}$ shows a zig-zag pattern rising from 32.576 mA/cm$^2$ to 32.929 mA/cm$^2$, where all other parameters are almost constant. This evident that even under those parametric sweeps, charge carriers can overwhelm intrinsic recombination losses within the BSF layer, because the carrier lifetime and diffusion length are large enough [19]. The optimum $WSe_2$ BSF parameters are taken considering both device performance and cost-effectiveness as thickness = 0.20 µm, acceptor concentration = $10^{18}$ cm$^{-3}$, and defect density = $10^{14}$ cm$^{-3}$. The device achieves $V_{OC}$ of 1.02 V, $J_{SC}$ of 32.472 mA/cm$^2$, FF of 88.212%, and PCE of 29.21% under these ideal circumstances.

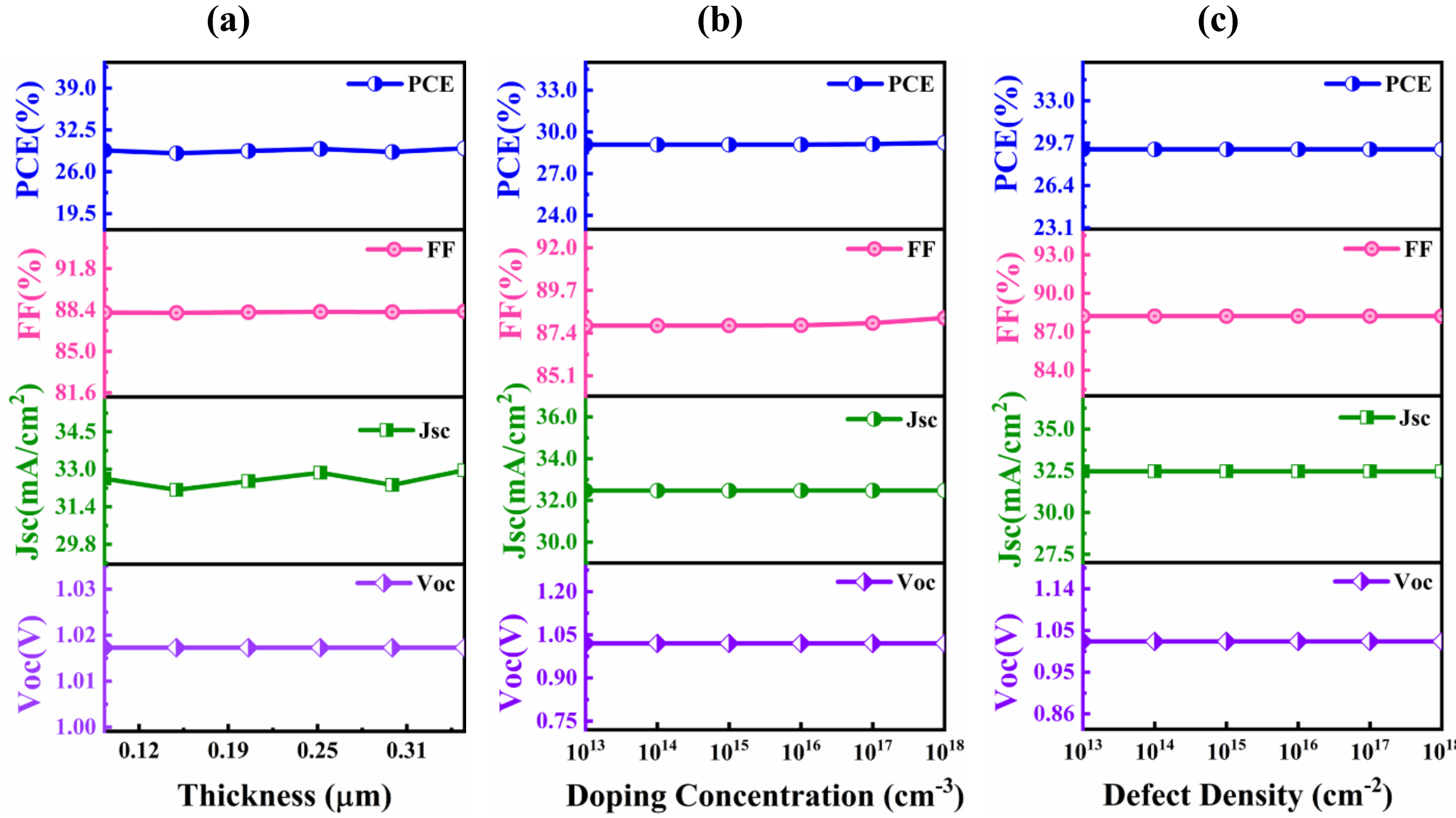

**Fig. 6:** Optimization of the performance parameters of SSCSe-based photonic device by changing (a) width (b) acceptor density and (c) bulk defects of the BSF.

### 3.5 *Impact of* Working Temperature *layer on PV performance of* the $SrScCu_3Se_4$ device

Temperature, Series resistance, and shunt resistance have a great influence on the performance of a PV solar cell. Fig. 7 shows the performance parameter of the SSCSe solar cell considering the variation of temperature, series resistance, and shunt resistance.

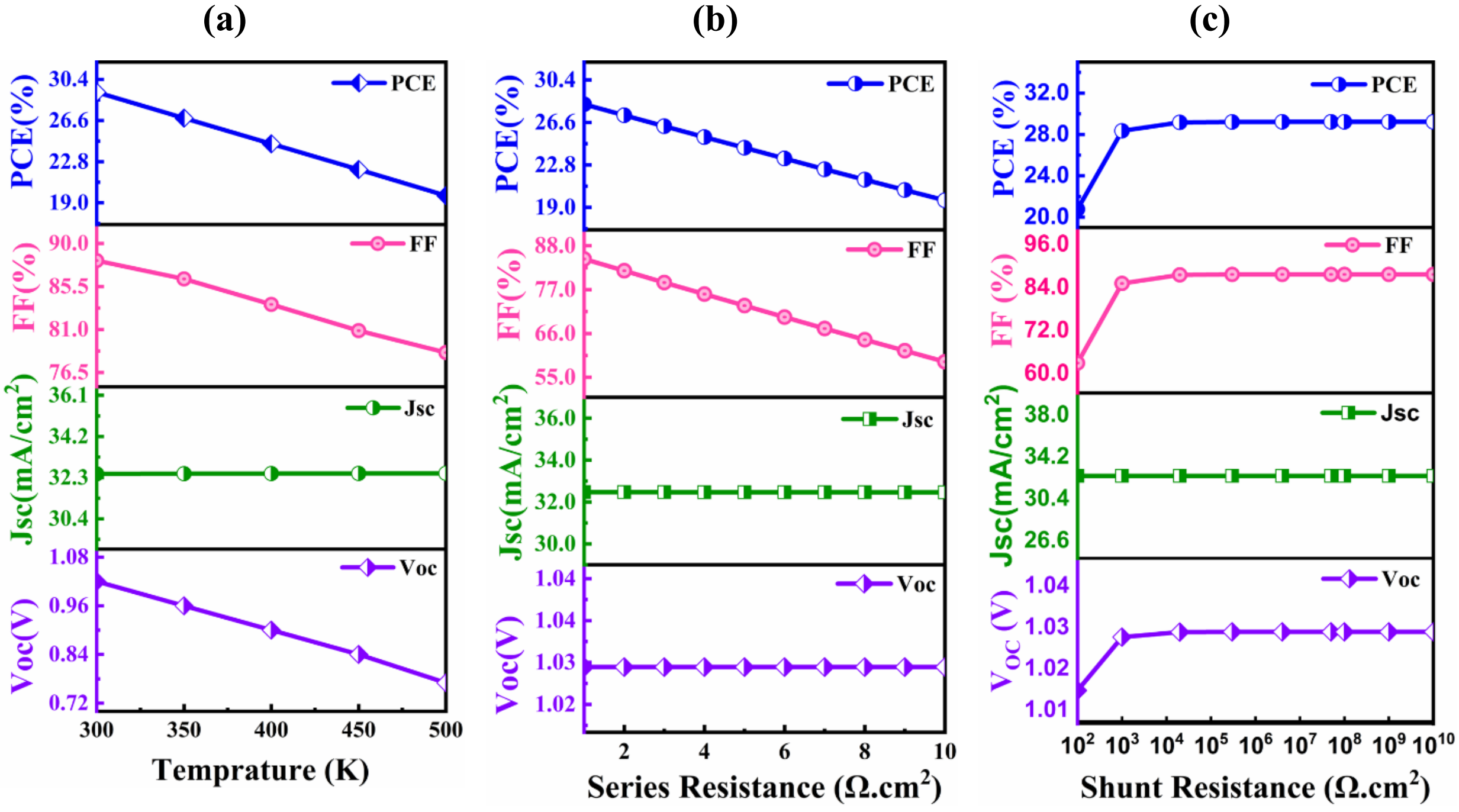

**Fig. 7:** The variation in PV metrics of the SSCSe-based photonic device as a function of absorber layer: (a) Working temperature, (b) Series resistance, and (c) Shunt resistance.

The performance of the PV solar cell is greatly affected by temperature. Fig. 7(a) shows that the performance of PV solar cell is significantly reduced as the operating temperature increases from 300 K to 500 K. $V_{OC}$ drops from 1.02 V to 0.77 V, FF lowers from 88.21% to 78.58%, and PCF drops from 29.21% to 19.66%. Conversely, $J_{SC}$ shows a little upward trend, increasing from 32.472 mA/cm$^2$ to 32.496 mA/cm$^2$. More thermal energy arises at higher temperatures, which improves electron-hole pair formation and slightly enhances the short circuit current; a minor improvement happens in JSC. From the $V_{OC}$–temperature equation, the temperature dependency of $E_g$ and $V_{OC}$ is typically accountable for the decline of $V_{OC}$ with increasing temperature [39].

$$\frac{dV_{oc}}{dT} = \frac{V_{OC}}{T} - \frac{E_g/q}{T} \quad (12)$$

According to Equation (12), as the term Eg/q is integrally greater than $V_{OC}$ and $V_{OC}$ is negatively affected by temperature increase, leading to its decrease. The reverse saturation current increases at narrower bandgap at higher temperatures [39]. Subsequently, the total performance of the solar cell weakens as both FF and PCE decrease.

However, $R_S$ should be close to zero, and $R_{Sh}$ should be an infinitely high value for an ideal solar cell. Shunt resistance is linked with leakage pathway and recombination produced by defects and device structural limitations, while series resistance comes from bulk resistance of materials and contact/interface resistance [38]. Increasing $R_S$ has a minor effect on $V_{OC}$, which remains almost constant at 1.02 V, as shown in Fig. 7(b). This is expected subsequently, in an open-circuit condition, $R_S$ has a negligible effect. However, resistance loss increases due to higher $R_S$, which try to stop carrier flow, resulting in JSC slightly dropping from 32.472 to 32.361 mA/cm$^2$ [40]. Subsequently, as RS increases from 1 to 10 Ω·cm$^2$, FF and PCE significantly decline, resulting weaken performance of the solar cell. In this case, PCE drops from 29.21% to 19.63% and FF drops from 87.63% to 58.91%.

Fig. 7(c) demonstrates the effect of varying $R_{Sh}$ from $10^2$ to $10^{10}$ Ω·cm$^2$. The performance fluctuation is very significant in this range. JSC remains almost constant at around 32.472 mA/cm$^2$, while $V_{OC}$, FF, and PCE exhibit a very notable rise though initially it shows lower value because of high sensitivity to $R_{Sh}$ at small values, where leakage paths are pronounced. This happens due to high $R_{Sh}$ is and the leakage current can be avoided compared to the photocurrent. Hence, lower $R_S$ and maintaining $R_{Sh}$ high are crucial for enhancing SSCSe solar cell efficiency.

### 3.6. Impact of $WSe_2$ Incorporation on $SrScCu_3Se_4$ Photonic Device Performance

The device output is highly sensitive to the electronic quality of the absorber because photogeneration occurs within this layer while defect states provide efficient pathways for non-radiative electron–hole recombination [41]. Consequently, recombination dynamics and carrier-collection efficiency strongly govern the photovoltaic response [42]. These effects are conveniently evaluated using the illuminated current density–voltage (J–V) characteristics, which relate the extracted current density to the applied terminal voltage and allow direct comparison of parameters such as $J_{SC}$ and Voc.

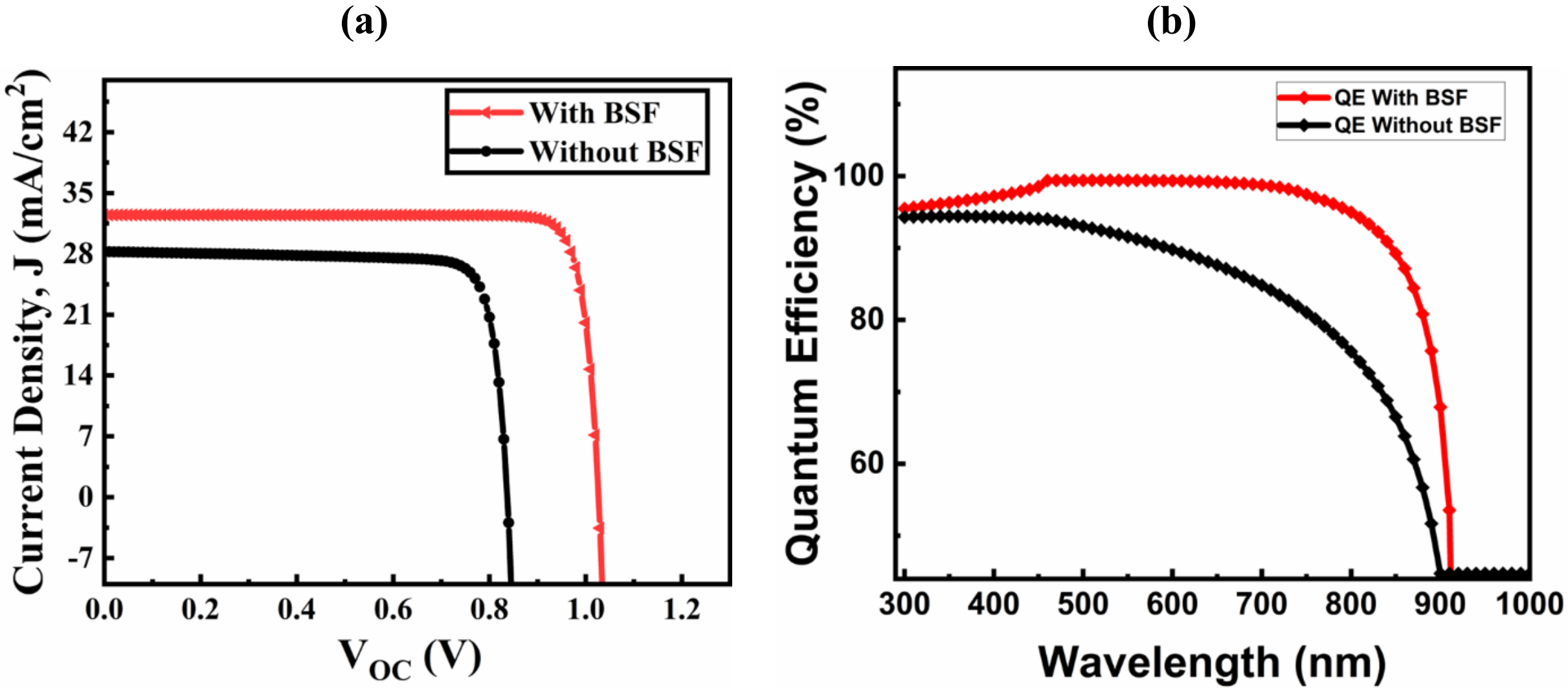

**Fig. 8:** Photovoltaic characteristics of the SSCSe photonic device, including (a) J–V curves, (b) quantum efficiency, are studied with and without the influence of the $WSe_2$ BSF layer

Fig. 8(a) shows the current vs. voltage curves of the SSCSe photonic device with and without $WSe_2$ layer. In the absence of the BSF, the current density remains close to 28.248 mA/cm$^2$ over the low-to-moderate voltage range and then approaches zero near an open-circuit voltage of approximately 0.84 V. After introducing the $WSe_2$ BSF, the device sustains a higher photocurrent density of about 32.472 mA/cm$^2$ and the J–V curve shifts to a larger open-circuit voltage, reaching roughly 1.03 V. The improvement is consistent with the role of a BSF in suppressing minority-carrier recombination at the rear interface and strengthening carrier selectivity near the back contact, which enhances carrier collection and reduces the dark (diode) recombination current [28]. As the applied voltage approaches Voc, the diode recombination current increases rapidly and progressively offset the photogenerated current, producing the sharp transition (knee) toward zero net current that is characteristic of photovoltaic diode behavior [43].

Fig. 8(b) presents the QE spectrum of the SSCSe device, demonstrating a pronounced wavelength dependence that reflects how effectively incident photons are converted into collected charge carriers. Without the BSF, the QE is about 94.2% at 300 nm and then decreases progressively beyond roughly 500 nm, indicating reduced carrier collection as the generation profile shifts deeper into the absorber and recombination losses become more influential. In contrast, when the $WSe_2$ BSF layer is introduced, the QE is 95.47% at 300 nm and rises rapidly with wavelength, reaching a broad maximum of approximately 99.4% across the 450–700 nm range. This enhanced response is consistent with improved photocarrier collection over the main absorption band, which can be attributed to the BSF-induced reduction of rear-interface recombination and the resulting increase in carrier selectivity near the back contact. The same improvement in collection is reflected in the higher current density observed in the J–V characteristics (Fig. 8(a)). Beyond 700 nm, the QE gradually decreases as the wavelength increases because the photon energy approaches the absorber band edge. In this region, optical absorption becomes weaker and fewer electron–hole pairs are generated. In addition, the carriers that are generated are more likely to recombine before they can be collected by the external circuit [28].

### 3.7 Thermal behavior of the proposed chalcogenide device

The thermal behavior of photovoltaic solar cell devices is a very critical aspect in both optical absorption and electrical performance, because the heat generated within the photovoltaic cell can vary transport characteristics, accelerate material and interface degradation over time, and rise recombination [44]. Only a small part of the absorbed photon energy under sunlight is converted into electrical power, while the rest of the energy is lost as heat through several channels [45]. In this work, the leading heat sources are Joule heating, which is mainly introduced from charge transport through areas with limited conductivity and contact/interface resistances, and nonradiative recombination heating, which is the result of SRH recombination. Discovering hot spots within the multilayer structure and recognizing weak areas that might require advanced passivation, defect control, or thermal management are made easier by investigating the spatial distribution of these heat-generation terms.

#### 3.7.1 Three-dimensional distribution of nonradiative recombination heating in the proposed PV device.

Fig. 9 displays three-dimensional maps of nonradiative recombination (SRH) heat generation for SSCSe-based solar cells under optimal bias, where the optimum voltage is 1.02 V, and short circuit, where the voltage is almost zero. Panels (a) show the same materials at V=0, and panels (b) represent SSCSe at V = 1.02 V. However, the heat generation is not uniform along all absorbers and is mostly found in the vicinity of the interface region and the junction of the two layers.

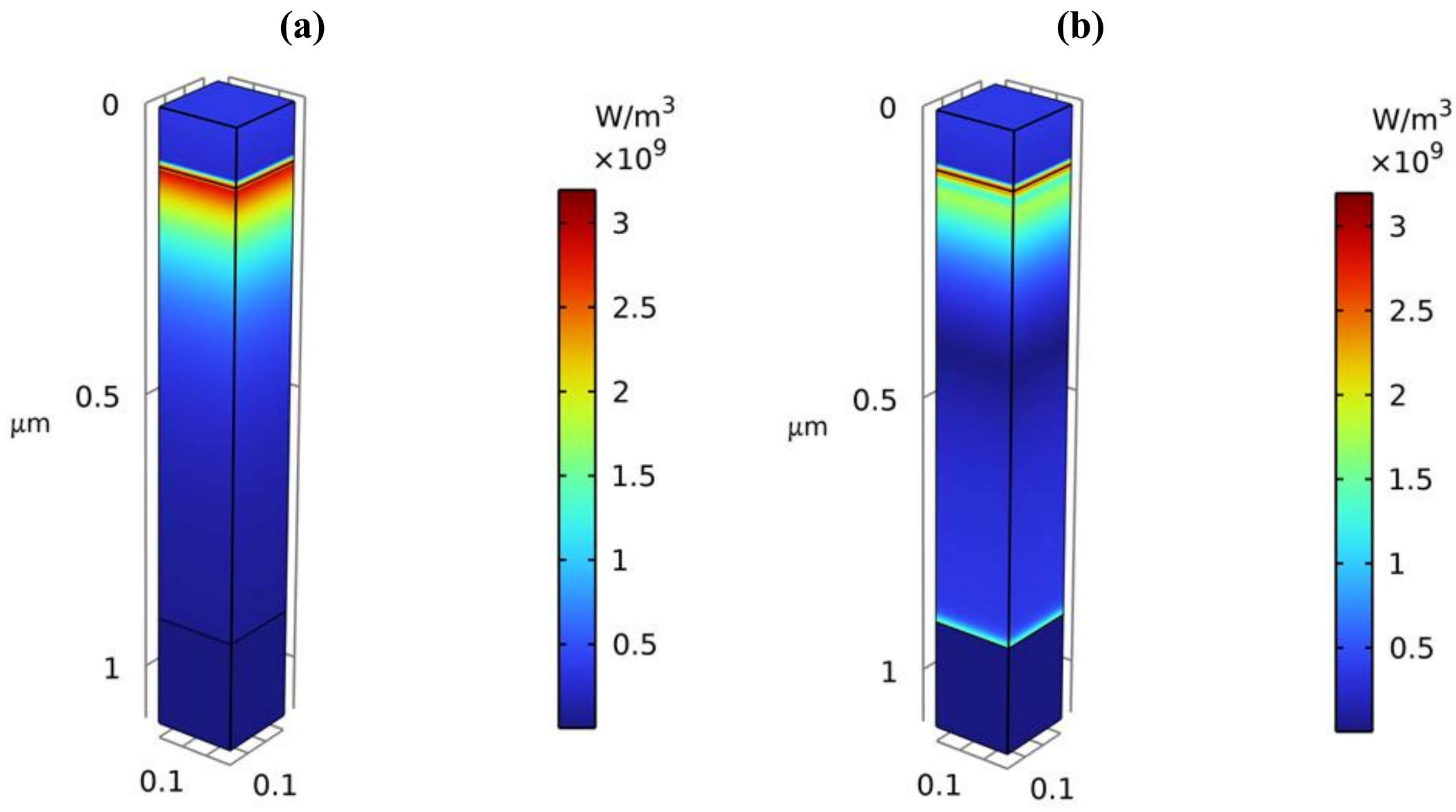

**Fig. 9:** Three-dimensional nonradiative recombination profiles ($W/m^3$) for SSCSe (a) At V=0 V (b) At V=1.02 V

From Fig. 9(a), it is observed that the absorber bulk region is less affected by heat at V = 0 V, and conversely, most of the SRH recombination heating is intensified near the ZnSe/absorber interface, which is the front heterojunction. For SSCSe, the highest heat-generation in this interface is approximately $1.64\times10^{9\cdot}$ $W/m^3$. Carriers are removed rapidly because of the internal electric field, and the steady-state carrier density in the absorber is decreased, which limits bulk trap-assisted recombination corresponding to forward bias. This behavior is observed under short-circuit operation [46].

At optimum operating voltage 1.02V, the nonradiative recombination heating is not restricted to a tiny interfacial region, as shown in Fig. 9(b). In fact, the maximum intensity is still centered close to the window-absorber heterojunction. Moderate heat generation is seen in the front section of the absorber layer. The maximum heat generation is $1.52\times10^9$ $W/m^3$ around the front contact. This result shows a developed steady-state carrier population under operating bias, so electron–hole overlap is enlarged, and SRH recombination is strengthened throughout areas where bulk traps exist [47]. Remarkably, SSCSe absorber also shows one more flashpoint for heat generation, and that is near the back interface junction, where the value of heat generation is around $1.09\times10^7$

W/m$^3$. This behavior indicates that interfacial recombination near the back contact is improved when forward bias is applied. Such characteristics can be described by carrier accumulation transported by disapproving band alignment or defective contact selectivity, which limits carrier abstraction and endorses local electron–hole existence at the edge [48].

### 3.7.2 Three-dimensional distribution of Joule recombination heating in the proposed PV devices.

Fig. 10 illustrates the three-dimensional Joule heating mode of the SSCSe-based PV solar cell at zero and optimal voltage. During carrier transport, resistive power dissipation is used by Joule heating and then scales with resistivity and current flow. Near operational points, Joule heat loss is less, which force to drop in current density, and is more obvious under short-circuit conditions, resulting the strongest photocurrent [26,49].

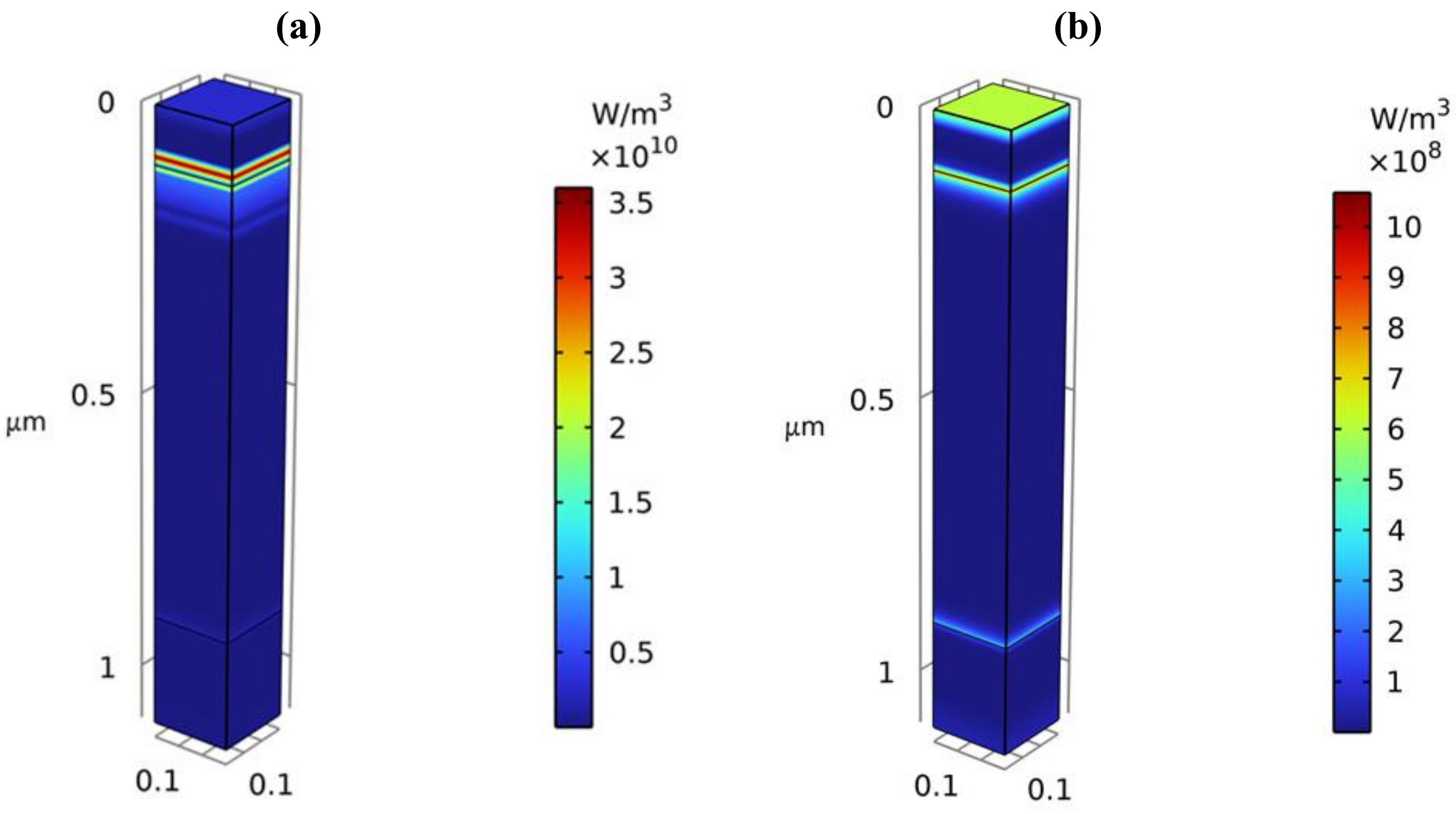

**Fig. 10:** Three-dimensional Joule heating profiles (W/m$^3$) for SSCSe (a) At V=0 (b) At V=1.02.

From Fig. 10 (a), the Joule heat is extremely non-uniform at V = 0 V and is intense at the front and rear surfaces and at both contacts, whereas the absorber bulk is less affected by heat. Due to transport obstacles, local electric-field development, and current crowding are typically strong enough in front of the contacts and electrode area, and that tends to more J·E dissipation than in the absorber, this surface and contact localization is very adequate [26,49]. For the SSCSe-based

PV solar cell, the maximum Joule-heating is found $2.75 \times 10^9$ W/m$^3$ at the window contact and $2.14 \times 10^8$ W/m$^3$ at the BSF contact, with larger values of $8.18 \times 10^9$ W/m$^3$ at the ZnSe/SSCSe junction contact and $6.82 \times 10^8$ W/m$^3$ at the SSCSe/$WSe_2$ junction contact. However, the short-circuit maps evident that Joule heat losses is a result of resistive dissipation in front of the junction and surface contact areas rather than by the absorber bulk.

Fig. 10(b) illustrates that, at operating voltage, Joule-heating shows a downward trend as compared to short circuit behavior for absorbers, ensuring that reduction of resistive dissipation occurs when the operating current density is less than at short circuit. For SSCSe, the loss magnitude falls to $6.05 \times 10^8$ W/m$^3$ (front contact), $4.41 \times 10^7$ W/m$^3$ (back contact), $6.8 \times 10^8$ W/m$^3$ (front interface), and $1.57 \times 10^8$ W/m$^3$ (rear interface). Although the Joule-heating level is much smaller at optimal voltage, the restraining dissipation is still intense in front of the surface contact and junction. As these areas generally exhibit the highest local electric fields and transport resistance, resulting $J \cdot E$ power loss [26,49].

### 3.7.3 Three-dimensional distribution of total Joule recombination heating in the proposed PV devices

Fig. 11 illustrates a three-dimensional profile of total heat generation for SSCSe-based solar cells under optimal bias and short circuit conditions. Heat generation is more pronounced in the top layers as the incident solar radiation is absorbed by them [45,50]. At the top surface when the voltage is zero, the total heat generated is $3.11 \times 10^9$ W/m$^{3,}$ and at the bottom surface, it is found as $2.11 \times 10^8$ W/m$^{3,}$ which is shown in Fig 11(a). Additionally, at the window-absorber junction, total heat generation is $2.06 \times 10^9$ W/m$^3$, where $2.11 \times 10^8$ W/m$^3$ occurs in the absorber-BSF junction. Similar trends are also seen in Fig. 11(b), where optimum voltage is applied. In the window-absorber junction, more heat is produced and is about $4.54 \times 10^8$ W/m$^{3,}$ and at the SSCSe/$WSe_2$ junction, it is $1.68 \times 10^8$ W/m$^3$. Near the front surface, the volumetric heat generation rate is $2.11 \times 10^8$ W/m$^{3,}$ and $2.11 \times 10^8$ W/m$^3$ for the back surface. This temperature distribution study is very important for heat distribution modeling of solar cells. This heat distribution is dynamic for ensuring both high efficiency and stability during operation.

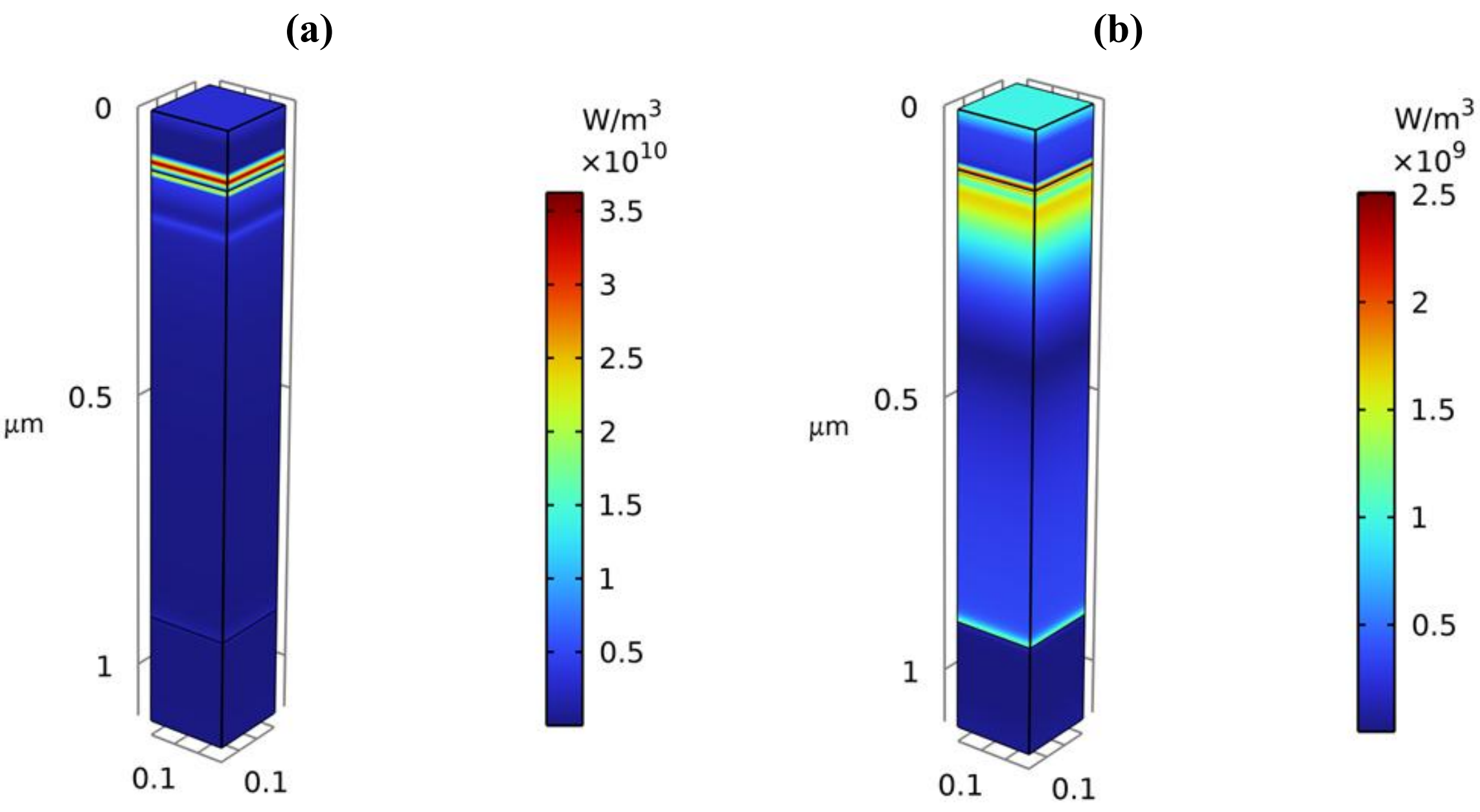

**Fig. 11:** Three-dimensional total heat source profiles (W/m$^3$) for SSCSe (a) At V=0 (b) At V=1.02.

### 4. Conclusion

The performance of three-dimensional SSCSe-based chalcogenide perovskite-inspired solar cells was simulated to evaluate their optical response, electrical output, and thermal loss characteristics. The simulation is done using a three-dimensional Multiphysics simulation tool, COMSOL Multiphysics (version 6.3). A polished, user-controlled mesh was used to ensure numerical stability and constant accuracy across the multilayer design. The device structure consists of the SSCSe absorber layer, a ZnSe window layer, and a $WSe_2$ BSF layer. By changing the absorber thickness, doping concentration, and defect density solar cell device was optimized while the window and BSF layers were fixed within the three-dimensional design. Under ideal situations, the device's optimal PCE, $V_{OC}$, $J_{SC}$, and FF were 29.217%, 1.02 V, 32.472 mA/cm$^2$, and 88.212%, respectively. Series resistance has a great impact on fill factor and PCE. Increasing series resistance mostly drops the FF and PCE, with a small effect on the $V_{OC}$ associated with parasitic resistance studies. Conversely, because of small leakage currents compared to the diode and photocurrent contributions**,** the effect of shunt resistance was seen as negligible once it exceeded approximately $10^4$ **Ω**.cm$^2$. QE analysis reflects the device efficiently transforms photons into charge carriers in the visible array, while absorption and carrier collection efficiency weaken in the near-infrared region. Finally Internal heat generation rate is mostly caused by nonradiative recombination and resistive dissipation joule heating, where the hotspots of heating develop close to material junctions, according to heat generation studies. Overall, the results show that SSCSe absorbers have suitable photovoltaic potential and demonstrate that the proposed three-dimensional modeling basis is a useful analytical tool for evaluating loss mechanisms and leading the experimental development of thermally stable, more efficient perovskite-inspired solar cells.

*Corresponding author: jak_apee@ru.ac.bd (Jaker Hossain)

**Conflicts of Interest**

The authors declare no conflicts of interest.

**Data Availability Statement**

The data that support the findings of this study are available from the corresponding author upon reasonable request.

**Declaration of generative AI and AI-assisted technologies**

The authors declare that no AI or AI-assisted tools were used in preparing this manuscript.